\documentclass[12pt]{article}
\pdfoutput=1

\usepackage{epsfig}
\usepackage{amsfonts}
\usepackage{amscd}
\usepackage{latexsym}
\usepackage{amsmath,amssymb}
\usepackage{verbatim}
\usepackage{setspace}
\usepackage{color}
\usepackage{fancyhdr}
\usepackage{cite}
\usepackage{hyperref}
\usepackage{tikz}
\usepackage{slashed}
\usetikzlibrary{calc}   
\usetikzlibrary{topaths}
\usetikzlibrary{decorations}
\usetikzlibrary{decorations.pathmorphing}

\usepackage{draft} 
\usepackage{hyperref}
\usepackage{graphicx,color,subfig}
\usepackage{cite}
\usepackage{mciteplus}
\usepackage{skak}
\usepackage[version=3]{mhchem}
\usepackage{empheq}
\newcommand*\widefbox[1]{\fbox{\hspace{0em}#1\hspace{0em}}}
\DeclareFontFamily{OT1}{pzc}{}
\DeclareFontShape{OT1}{pzc}{m}{it}{<-> s * [1.10] pzcmi7t}{}
\DeclareMathAlphabet{\mathpzc}{OT1}{pzc}{m}{it}

\numberwithin{equation}{section}

\def\e{{\varepsilon}}

 \def\bz{{\bar z}}
 
\def\0{{(0)}}
\def\1{{(1)}}
\def\2{{(2)}}

\def\<{\langle }
\def\>{\rangle }

\def\eads${H$_3$}

\newcommand{\ba}{\begin{align}}
\newcommand{\ea}{\end{align}}
\def\be{\begin{equation}}
\def\ee{\end{equation}}
\def\beq{\be\begin{array}{c}}
\def\eeq{\end{array}\ee}

\def\be#1\ee{\begin{align}#1\end{align}}

\begin{document}

\newcommand{\scriplus}{\mathcal{I}^+}

\newcommand{\Del}{\nabla}

	 \renewcommand{\theequation}{\thesection.\arabic{equation}}
   \makeatletter
  \let\over=\@@over \let\overwithdelims=\@@overwithdelims
  \let\atop=\@@atop \let\atopwithdelims=\@@atopwithdelims
  \let\above=\@@above \let\abovewithdelims=\@@abovewithdelims
\renewcommand\section{\@startsection {section}{1}{\z@}%
                                   {-3.5ex \@plus -1ex \@minus -.2ex}
                                   {2.3ex \@plus.2ex}%
                                   {\normalfont\large\bfseries}}

\renewcommand\subsection{\@startsection{subsection}{2}{\z@}%
                                     {-3.25ex\@plus -1ex \@minus -.2ex}%
                                     {1.5ex \@plus .2ex}%
                                     {\normalfont\bfseries}}

\renewcommand{\H}{\mathcal{H}}
\newcommand{\SU}{\mbox{SU}}
\newcommand{\chiu}{\chi^{{\rm U}(\infty)}}
\newcommand{\ff}{\rm f}
\linespread{1.3}

\unitlength = .8mm

\begin{titlepage}

\begin{center}

\hfill \\
\hfill \\
\vskip 1cm

\title{Gluon Amplitudes as $2d$ Conformal Correlators}

\author{Sabrina Pasterski,$^1$ Shu-Heng Shao,$^2$ and Andrew Strominger$^1$}

\address{
$^1$Center for the Fundamental Laws of Nature, Harvard University,\\
Cambridge, MA 02138, USA
\\
$^2$School of Natural Sciences, Institute for Advanced Study, \\Princeton, NJ 08540, USA
}

\end{center}

\vspace{2.0cm}

\begin{abstract}

Recently, spin-one wavefunctions in  four dimensions that are conformal primaries of the Lorentz group $SL(2,\mathbb{C})$  were constructed.  We compute low-point, tree-level  gluon scattering amplitudes in the space of these conformal primary wavefunctions.  The answers have the same conformal covariance as correlators of spin-one primaries in a $2d$ CFT.  The BCFW recursion  relation between three- and four-point gluon amplitudes is recast into this conformal basis.

\end{abstract}

\vfill

\end{titlepage}

\eject
\tableofcontents
~\\

\section{Introduction}

The $4d$ Lorentz group acts as the global $SL(2,\mathbb{C})$ conformal group on the celestial sphere at null infinity. This implies that the 
$4d$ massless quantum field theory (QFT) scattering amplitudes, recast as correlators 
on the celestial sphere, share some properties with those of $2d$ 
CFTs. When gravity is included, the plot thickens: the global conformal group is enhanced to the infinite-dimensional local group \cite{Cachazo:2014fwa,Kapec:2014opa,Cheung:2016iub,Kapec:2016jld},\footnote{Up to IR divergences at one loop \cite{Bern:2014oka,He:2014bga,Cachazo:2014dia,He:2017fsb,Kapec:2017tkm}.} suggesting an even tighter connection with $2d$ CFT.  This 
$4d$-$2d$ connection is of interest both for the ambitious goal  
of a holographic reformulation of flat space quantum gravity, as well as for its potentially strong yet unexploited mathematical implications for the rich subject of QFT scattering  amplitudes.

A perhaps-not-too-distant goal is to find out whether or to what extent there is any set of QFT scattering amplitudes that can be approximately\footnote{Exact CFTs are expected only with the inclusion of gravity. What we have in mind here is something like a large $N$ approximation.} generated by some kind of $2d$ CFT. 
It has been clear from the outset that such CFTs would not be of the garden variety with operators in highest weight representations. Recently it has emerged \cite{Pasterski:2017kqt} from the study of two-point functions,  that the unitary principal continuous series (which has appeared in a variety of  CFT studies \cite{Costa:2012cb,Gadde:2017sjg,Hogervorst:2017sfd,Caron-Huot:2017vep}) of the Lorentz group plays a central role. Beyond that it is not clear what to expect, and concrete computations are in order. 
In this paper we compute and explore some basic properties of amplitudes - in particular the three- and four-point tree-level gluon  amplitudes - presented as $2d$ conformal correlators on the celestial sphere. This case is of special interest both because it has been shown that soft gluons generate a $2d$ current algebra  \cite{Strominger:2013lka,He:2015zea} and because of the plethora of beautiful results about these amplitudes in the momentum-space representation. In principle, one may hope to find a $2d$ CFT 
of some kind that generates the $4d$ tree amplitudes. Indeed several  constructions including the scattering equations \cite{Cachazo:2013hca} and  twistor space \cite{Witten:2003nn,ArkaniHamed:2009si,Mason:2009sa}  seem close to achieving this goal.

The $SL(2,\mathbb{C})$ conformal presentation of $4d$ QFT scattering amplitudes was discussed long ago by Dirac \cite{Dirac:1936fq}. Recently, soft photon and gluon theorems were recast as $2d$ Kac-Moody-Ward identities \cite{Strominger:2013lka,He:2015zea,Lipstein:2015rxa,Cardona:2015woa,Nande:2017dba}. 
Massless \cite{Cheung:2016iub} and massive \cite{Pasterski:2016qvg} scalar wavefunctions in four-dimensional Minkowski spacetime that are primaries of the Lorentz group $SL(2,\mathbb{C})$ were constructed.  Such solutions, called  \textit{conformal primary wavefunctions}, are labeled by a point $z,\bar z$ in $\mathbb{R}^2$ and a conformal dimension $\Delta$, rather than the three independent components of an on-shell four-momentum.  The massless spin-one conformal primary wavefunctions in four dimensions were also constructed  in \cite{Cheung:2016iub}.  A comprehensive survey of  conformal primary wavefunctions with or without spin in arbitrary spacetime dimensions was performed in \cite{Pasterski:2017kqt}. In particular, it was shown there that conformal primary wavefunctions in $\mathbb{R}^{1,d+1}$ with conformal dimensions on the principal continuous series $\Delta\in \frac d2+i\mathbb{R}$ of $SO(1,d+1)$ form a complete set of delta-function-normalizable solutions to the wave equation. The factorization singularity of  amplitudes in the conformal basis is studied in \cite{Cardona:2017keg,NVWZ}. 

In this paper we  study low-point $4d$  tree-level gluon scattering amplitudes in the space of conformal primary wavefunctions.  We find that the tree-level color-ordered MHV four-point amplitude $1^- 2^- \to3^+4^+$ takes the form
\begin{align}\label{drf}
\mathcal{\tilde A}_{--++} (z_i,\bar z_i) =  I(z_{ij},\bar z_{ij}) \, \delta(\sum_{i=1}^4 \lambda_i)\delta(|z-\bz|) {z^{5\over3}\over (z-1)^{1\over3}}\,,~~~~~~1<z,
\end{align}
where $z,\bar z$ are the cross ratios, $I(z_{ij},\bar z_{ij})$ is a product of powers of $z_{ij},\bz_{ij}$ that is fixed by conformal covariance, and  $1+i\lambda_i$ are the conformal dimensions of the four primaries. The delta-function for the imaginary part of the cross ratio is shown to be implied by $4d$ translation invariance (which is generally obscured in the conformal basis).  As an important check on this formula, we show that is has a BCFW representation in a factorization channel  involving the product of three-point functions.

Going forward, more might be learned from a $2d$ conformal block expansion of the four-point function \eqref{drf}. Interestingly, the BCFW relation resembles the OPE expansion of a $2d$ CFT four-point correlator. We leave these directions  to future work.

The paper is organized as follows. In Section \ref{sec:2} we review  the massless vector conformal primary wavefunction in four spacetime dimensions. 
 The change of basis from plane waves to the conformal primary wavefunctions is implemented by a Mellin transform. 
 In Sections \ref{sec:MHV3pt} and \ref{sec:MHV4pt}, we compute Mellin transforms of the tree-level three- and four-point amplitudes, and show that the answers transform as spin-one $2d$ conformal correlators.  In Section \ref{sec:BCFW}, we write the BCFW recursion  relation for the MHV four-point amplitude in the  conformal primary wavefunction basis.  In Appendix \ref{app:convention} we set up our conventions for the spinor helicity variables.  In Appendix \ref{sec:2pt}, we review  Mellin transforms on  inner products of two gauge boson one-particle  states.

\section{A Conformal Basis for Gauge Bosons}\label{sec:2}

In this section we review the massless vector conformal primary wavefunctions  in $\mathbb{R}^{1,3}$ \cite{Cheung:2016iub,Pasterski:2017kqt}, i.e. solutions to the  four-dimensional Maxwell equation that transform as two-dimensional spin-one conformal primaries.  In particular, the transition from momentum space to conformal primary wavefunctions is implemented by a Mellin transform.  Later sections  will derive and study  low-point  amplitudes in the new conformal basis. 

Let us begin by setting up the notations.  We will restrict ourselves to four-dimensional Minkowski space $\mathbb{R}^{1,3}$ with spacetime coordinates $X^\mu$ ($\mu=0,1,2,3)$.  We use $z, \bar z$ to denote a point in $\mathbb{R}^2$. We will sometimes use $\partial_+$ $(\partial_-)$ to denote $\partial_z$ $(\partial_{\bar z})$.
\subsection{Massless Vector Conformal Primary Wavefunctions}\label{sec:CPW}
Scattering problems of gauge bosons are usually studied in the plane wave basis which, in Lorenz gauge $\partial^\mu A_\mu=0$,  consists of
\begin{align}\label{planewave}
\epsilon_{\mu\ell}(p) \, e^{\mp i | \vec p| X^0  \pm i \vec p \cdot \vec X}\,,~~~~~\ell =\pm1\,.
\end{align}
Here $\epsilon_{\mu\ell}(p)$ are the polarization vectors for the helicity $\ell$ one-particle states. They  satisfy $\epsilon_\ell(p) \cdot p=0$, $\epsilon_\pm(p)^* =\epsilon_{\mp}(p)$, and $\epsilon_\ell(p) \cdot \epsilon_{\ell'}(p)^* =\delta_{\ell \ell'}$. 
Plane wave solutions are labeled by three continuous variables and two signs:
\begin{itemize}
\item a spatial momentum $\vec p$,
\item a $4d$ helicity $\ell=\pm1$,
\item a sign distinguishing an incoming solution from an outgoing one.
\end{itemize}

To make the two-dimensional conformal symmetry  manifest,  an alternative set of solutions for the Maxwell equation, called \textit{massless vector conformal primary wavefunctions}  
\begin{align}
V^{\Delta \pm}_{\mu J}(X^\mu;z,\bar z)\,,
\end{align}
 were constructed in \cite{Cheung:2016iub,Pasterski:2017kqt}.
  These solutions are again labeled by   three continuous variables and two signs:
\begin{itemize}
 \item a point $z,\bar z$ in $\mathbb{R}^2$ and a conformal dimension $\Delta$,
 \item  a $2d$ spin $J=\pm1$,
 \item a sign distinguishing an incoming solution from an outgoing one. 
 \end{itemize}
The defining properties of the massless vector conformal primary wavefunction  $V^{\Delta\pm}_{\mu J}(X^\mu;z,\bar z)$ are:
\begin{enumerate}
\item It satisfies the Maxwell equation,
\begin{align}
\left(  {\partial\over \partial X^\rho} {\partial\over \partial X_\rho} \delta_{\nu }^\mu   - {\partial\over \partial X^\nu}{\partial\over \partial X_\mu}\right)    V^{\Delta \pm}_{\mu J} (X^\mu;z,\bar z) =0\,.
\end{align}
\item It transforms as a four-dimensional vector  and a two-dimensional (quasi-)conformal primary with spin $J=\pm1$ and has dimension $\Delta$  under an $SL(2,\mathbb{C})$ Lorentz transformation:
\begin{align}\label{covarianceV}
&V^{\Delta \pm}_{\mu J } \left(\Lambda^\mu_{~\nu} X^\nu ; {az+b\over cz+d} , {\bar a \bar z + \bar b \over \bar c \bar z +\bar d}\right)
= (cz+d)^{\Delta+ J }  (\bar c\bar z +\bar d)^{\Delta- J}\, \Lambda_\mu ^{~\rho} V^{\Delta\pm}_{\rho J}(X^\mu;z,\bar z)\,,
\end{align}
where $a,b,c,d\in \mathbb{C}$ with $ad-bc=1$ and $\Lambda^\mu_{~\nu}$ is the associated $SL(2,\mathbb{C})$ group element in the four-dimensional representation.\footnote{For an explicit expression for $\Lambda^\mu_{~\nu}$ in terms of $a,b,c,d$, see \cite{Pasterski:2016qvg}.} 
\end{enumerate}
By construction, scattering amplitudes of conformal primary wavefunctions  transform covariantly as a two-dimensional conformal correlators of spin-one primaries with conformal dimensions $\Delta_i$.

Let us write down the explicit expression for the  massless vector conformal primary wavefunctions.   We define a ``unit" null vector $q^\mu$ associated to $z,\bar z$ as
\begin{align}\label{fga}
q^\mu (z, \bar z)  = ( 1+|z|^2 , z+\bar z , -i (z-\bar z) , 1-|z|^2)\,.
\end{align}
Under an $SL(2,\mathbb{C})$ transformation $z\to z'=(az+b)  /(cz+d), \bar z\to \bar z'=(\bar a\bar z+\bar b)  /(\bar c\bar  z+\bar d)$, the null vector $q^\mu$ transforms as a vector up to a conformal weight,
\begin{align}
q^\mu \to{ q^\mu }'   =  (cz+d)^{-1}(\bar c \bar z+\bar d)^{-1}  \Lambda^\mu_{~\nu} q^\nu\,.
\end{align}
The derivative of $q^\mu$ with respect to $z$ and $\bar z$ are respectively the polarization vectors of helicity $+1$ and $-1$ one-particle states propagating in the $q^\mu$ direction:
\begin{align}\label{polarization}
\partial_z q^\mu =\sqrt{2} \, \epsilon_+^{\mu}(q)= (\bar z, 1, -i  , - \bar z)\,,~~~~~~
\partial_{\bar z}q^\mu  =\sqrt{2} \, \epsilon_-^{\mu}(q)= ( z, 1 , i , -z)\,.
\end{align}
They satisfy $q\cdot \partial_z q=q\cdot \partial_{\bar z} q= 0$, $\partial_z q \cdot \partial_z q = \partial_{\bar z} q\cdot \partial_{\bar z} q = 0$ and $\partial_z q \cdot \partial_{\bar z}q =2$. 

The explicit expression for the conformal primary wavefunctions was given in terms of the spin-one bulk-to-boundary propagator in the three-dimensional hyperbolic space $H_3$ in \cite{Cheung:2016iub,Pasterski:2017kqt}.  To compute gauge invariant physical observables such as scattering amplitudes, we can choose a convenient gauge representative for the conformal primary wavefunction.  In \cite{Cheung:2016iub,Pasterski:2017kqt} it was shown that, for $\Delta\neq1$, the vector conformal primary wavefunction  is gauge equivalent to 
\begin{align}\label{gaugerep}
V_{\mu J}^{\Delta,\pm} (X^\mu; z, \bar z)= \mathcal{N} {\partial_Jq_\mu\over (-q\cdot X\mp i\epsilon)^\Delta}\,,~~~\Delta\neq 1\,,
\end{align}
where $\mathcal{N}=  { (\mp i)^\Delta \Gamma(\Delta)/\sqrt{2}} $ is a normalization constant chosen for later convenience.  From now on we will assume $\Delta\neq 1$.  (When $\Delta=1$, the conformal primary wavefunction itself is a total derivative in $X^\mu$, so is a pure gauge.  We have put in an $i\epsilon$-prescription to circumvent the singularity at the light sheet where $q\cdot X=0$.)  There is another set of conformal primary wavefunctions that are shadow to \eqref{gaugerep} \cite{Pasterski:2017kqt}.   We leave the study of scattering amplitudes in the shadow basis for future investigation.

Finally, we need to determine the range of the conformal dimension $\Delta$. It was shown in \cite{Pasterski:2017kqt} that the conformal primary wavefunctions are a complete and delta-function-normalizable basis if $\Delta$ ranges over  the one-dimensional locus 
\begin{align}
\mathcal{C} =\{\Delta\in 1+ i \mathbb{R}\}\,,
\end{align}
on the complex plane.   See also \cite{deBoer:2003vf,Cheung:2016iub} for an alternative argument via the hyperbolic slicing of Minkowski space.
For a given spin,   this range of $\Delta$ is known as the principal continuous series of  unitary representations of $SL(2,\mathbb{C})$.

\begin{table}
\centering
\begin{tabular}{|c|c|c|c|}
\hline
Bases & Plane Waves 
& Conformal Primary Wavefunctions \\
\hline
\text{Notations}&$\,  \epsilon_{ \mu \ell }(p)\exp\left[ \mp i |\vec p| X^0 \pm i \vec p \cdot \vec X\right]\,$  
&$V^{\Delta , \pm }_{\mu J}(X;z,\bar z)$ \\
\hline
Continuous Labels & $\vec p$&$ \Delta , z,\bar z$\\
\hline
Discrete Labels &$4d$ Helicity $\ell$ & $2d$ Spin $J$\\
& Incoming vs. Outgoing &  Incoming vs. Outgoing\\
\hline
\end{tabular}
\caption{A comparison between the plane waves and the massless vector conformal primary wavefunctions. The two set of solutions are labeled by some continuous labels and discrete labels.  The continuous labels for the former are  a spatial momentum $\vec p$, while those for the latter are $\Delta, z,\bar z$.  The  discrete labels consist of a sign distinguishing between  incoming and outgoing wavefunctions, and a sign for the $4d$ helicity  or the $2d$ spin. }
\end{table}

\subsection{Mellin Transform}\label{sec:Mellin}

The gauge representative \eqref{gaugerep} for the conformal primary wavefunction has the advantage that it is simply related to the plane wave \eqref{planewave} by a Mellin transform\footnote{The Mellin transform of a function $f(\omega)$ is defined by
\begin{align}
\widetilde f(\Delta) = \int_0^\infty d\omega \, \omega^{\Delta-1} f(\omega)\,,
\end{align}
while the inverse Mellin transform of $\tilde f(\Delta)$ is
\begin{align}\label{mellin}
f(\omega) = {1\over 2\pi i} \int_{c-i\infty}^{c+i\infty} d\Delta \, \omega^{-\Delta} \widetilde f(\Delta)\,,~~~~~c\in \mathbb{R}\,.
\end{align}}
\begin{align}
\mathcal{N} \,  {\partial_J q_\mu\over( -q\cdot X \mp i\epsilon)^\Delta}=
{  \partial_J q_\mu
\over{\sqrt{2}}}\, \int_0^\infty d\omega \, \omega^{\Delta-1} \, 
e^{
\pm i \omega q\cdot X  -  \epsilon\omega }\,,~~~~\Delta=1+i\lambda\,,
\end{align}
with the plus (minus) sign for an outgoing (incoming) wavefunction. 
$\Delta$ is the scaling dimension under the Lorentz boosts which preserve the particle trajectory while rescaling $q\cdot X$.   

Notice that the spin $+1 (-1)$, outgoing conformal primary wavefunction $A^{\Delta, +}_{\mu +}$ ($A^{\Delta, +}_{\mu -}$) has nontrivial projections to the $4d$ helicity $+1 (-1)$ sector. It follows that a $4d$ helicity $+1$ $(-1)$ \textit{outgoing} one-particle state is mapped to a $2d$ spin $+1$ ($-1$) conformal operator under Mellin transform.  By CPT, a $4d$ helicity $+1$ $(-1)$ \textit{incoming} one-particle state is mapped to a $2d$ spin $-1$ ($+1$) conformal operator under Mellin transform.

When considering scattering amplitudes in the plane wave basis, it is often convenient to take all particles to be outgoing but some of them carrying negative energy. In  the basis of conformal primary wavefunctions, however, we need to keep track of  which particles are incoming and outgoing, and different crossing channels have to be treated separately. 
Nonetheless, we will label the helicity of an external gauge boson as if  it were an outgoing particle.

This change of basis can be immediately extended to any gauge boson scattering amplitude.  Consider a general $n$-point gauge boson scattering amplitude, which is a function of the $\omega_i,z_i,\bar z_i$ and helicities $\ell_i$,\begin{align}
\mathcal{A}_{\ell_1\cdots\ell_n} (\omega_i ,z_i ,\bar z_i)\,,
\end{align}
with the momentum conservation delta function $\delta^{(4) }(\sum_{i=1}^n p^\mu_i)$ included.
We can perform a Mellin transform on each of the external particles to go to the basis of conformal primary wavefunctions,
\begin{align}\label{eq:mt}
\mathcal{\tilde A}_{J_1\cdots J_n}(\lambda_j, z_j,\bar z_j)
=  \prod_{i=1}^n  \int_0^\infty d\omega_i  \, \omega_i^{\lambda_i } \,
\mathcal{A}_{\ell_1\cdots \ell_n} (\omega_j,z_j,\bar z_j)\,,
\end{align}
where the $2d$ spin $J_i$ is identified as  the $4d$ helicity $\ell_i$, i.e. $J_i  = \ell_i$.
From the defining properties of the conformal primary wavefunction \eqref{covarianceV}, the resulting function $\mathcal{\tilde A}_{J_1\cdots J_n}(\lambda_i,z_i,\bar z_i)$   is guaranteed to transform covariantly as a two-dimensional conformal correlators of spin-one primaries with dimensions $\Delta_i=1+i\lambda_i$, i.e.
\begin{align}\label{conformalcov}
\mathcal{\tilde A}_{J_1\cdots J_n}\left(\lambda_j,
{a z_j+b\over cz_j +d} ,{\bar a \bar z_j +\bar b\over\bar c\bar z_j+\bar d}\right)
  =  \prod_{i=1}^n \left[  
  (cz_i +d)^{\Delta_i  + J_i}
   (\bar c \bar z_i+\bar d)^{\Delta_i - J_i}  \right]
   \mathcal{\tilde A}_{J_1\cdots J_n}(\lambda_j,z_j,\bar z_j)\,.
\end{align}

We will sometimes  use the inverse Mellin transform to convert an amplitude in the conformal primary wavefunction basis back to the plane wave basis,
\begin{align}
\mathcal{A}_{\ell_1\cdots\ell_n} (\omega_j , z_j ,\bar z_j)  =\prod_{i=1}^n  \int_{-\infty}^\infty  {d\lambda_i\over 2\pi}  \,\omega^{-1-i\lambda_i} \, \mathcal{\tilde A} _{J_1\cdots J_n}(\lambda_j,z_j,\bar z_j)\,.
\end{align}

To sum up the above discussion, the change of  basis from  plane waves to the massless vector conformal primary wavefunction is implemented by a Mellin transform \eqref{mellin} with conformal dimension $\Delta =1+i\lambda$ lying on the principal continuous series.  The $4d$ helicity of an  one-particle state is identified as the  $2d$ spin for the primary operator.

We now discuss some general properties of Mellin transforms of  low-point gluon amplitudes. They scale homogeneously under uniform rescaling of the frequencies as  
\begin{align}\label{eq:sc}
\mathcal{A}_{\ell_1\cdots\ell_n} (\Lambda\omega_i ,z_i ,\bar z_i)\,=\Lambda^{-n} \mathcal{A}_{\ell_1\cdots\ell_n} (\omega_i ,z_i ,\bar z_i)\,.
\end{align}
The momentum conservation delta function is included in $\mathcal{A}$ and contributes a factor of $\Lambda^{-4}$ in \eqref{eq:sc}. (Note this scaling also holds true for tree-level $\phi^4$ theory, so that the analysis in the rest of this subsection carries over to that case as well.)
It is convenient to  change the integration variables to an overall frequency $s\equiv\sum_i \omega_i$ and a set of ``simplex variables" $\sigma_i\equiv s^{-1}\omega_i\in[0,1]$  with $\sum_{i=1}^n\sigma_n=1$,
\begin{align}
\prod_{i=1}^n  \int_0^\infty d\omega_i  \, \omega_i^{i\lambda_i } [...]= \int_0^\infty ds s^{n-1+i\sum\limits_i\lambda_i}\prod_{i=1}^n  \int_0^1 d\sigma_i  \, \sigma_i^{\lambda_i }\delta(\sum\limits_i\sigma_i-1) [...]~.
\end{align}
Let  $A_{\ell_1\cdots \ell_n}$ denote the stripped amplitude with the delta function and an overall power of $s$ factored out:
\be
\mathcal{A}_{\ell_1\cdots \ell_n} (\omega_j,z_j,\bar z_j)= s^{-n}\,A_{\ell_1\cdots \ell_n} (\sigma_j,z_j,\bar z_j)\delta^{(4)}(\sum\limits_i\e_i\sigma_iq_i)\,,
\ee
where $q_i(z_i,\bz_i)$ is given  in~(\ref{fga}) (or its $(-+-+)$ signature analog which will be discussed in Section~\ref{sec:MHV3pt}) and $\e_i=+1$ $(-1)$ for an outgoing (incoming) external particle. 
Then, using~(\ref{eq:sc}) and
\begin{align}\label{mellindelta}
 \int_0^\infty d\omega \,\omega^{i\lambda-1} 
 =2\pi \,  \delta(\lambda)\,,
\end{align}
 we can rewrite the amplitude  in the conformal basis (\ref{eq:mt}) as:
\begin{align}\label{eq:mts}
\mathcal{\tilde A}_{J_1\cdots J_n}(\lambda_j, z_j,\bar z_j)
=2\pi\delta(\sum\limits_i\lambda_i)  \prod_{i=1}^n  \int_0^1 d\sigma_i  \, \sigma_i^{i\lambda_i } \,
A_{\ell_1\cdots \ell_n} (\sigma_j,z_j,\bar z_j)\delta^{(4)}(\sum\limits_i\e_i\sigma_iq_i)\delta(\sum\limits_i\sigma_i-1)\,.
\end{align}
    We thus have a total of 5 delta functions inside the integral of~(\ref{eq:mts}), which  localize the $\sigma_i$ integrals for up to five-point scattering amplitudes.  

We  can then compute $\mathcal{\tilde A}_{J_1\cdots J_n}(\lambda_j, z_j,\bar z_j)$ for $n\le 5$ easily in two steps.  In the first step, we rewrite the delta functions as 
\begin{align}
\delta^{(4)}(\sum\limits_i\e_i\sigma_iq_i)\delta(\sum\limits_i\sigma_i-1)
=
 C(z_i,\bar z_i) \prod_{i=1}^{n\le 5} \delta(\sigma_i-\sigma_{*i})\,,
 \end{align}
 for some function $C(z_i,\bar z_i)$. 
Here $\sigma_{*i}(z_j,\bar z_j)$'s are the solutions of $\sigma_i$'s fixed by the momentum conservation delta functions. For $n=3,4$, the momentum conservation equations are over-constraining, and the function $C(z_i,\bar z_i)$ will contain delta functions in $z_i,\bar z_i$, restricting the angles of the external particle trajectories. For example, as we will see  in Section \ref{sec:MHV4pt}, for $n=4$, $C(z_i,\bar z_i) \sim \delta(|z-\bar z|)$ where $z,\bar z$ are the cross ratios.

 In the second step, all the simplex integrals in $\sigma_i$ can be done by simply evaluating the integrand at $\sigma_i=\sigma_{*i}(z_j,\bar z_j)$.  The final result for the Mellin transform of the amplitude (\ref{eq:mts}) is then:
  \begin{align}\label{eq:mtsa}
\mathcal{\tilde A}_{J_1\cdots J_n}(\lambda_j, z_j,\bar z_j)
=2\pi\delta(\sum\limits_i\lambda_i)  \left(\prod_{i=1}^{n\le 5}   \sigma_{*i}^{i\lambda_i }\right) \,
A_{\ell_1\cdots \ell_n} (\sigma_{*j},z_j,\bar z_j)C(z_i,\bar z_i) \prod_{i}\mathbf{1}_{[0,1]}(\sigma_{*i})\,,
\end{align}
where the indicator function $\mathbf{1}_{[0,1]}(x)$  defined as
 \begin{align}\label{indicator}
 \mathbf{1}_{[0,1]}(x) =
 \begin{cases}
 1\,,~~~~~\text{if}~~x\in [0,1]\,,\\
 0\,,~~~~~\text{otherwise}\,,
 \end{cases}
 \end{align}
comes from the restricted range of $\sigma_i$ between 0 and 1 in their definitions.  
In Sections \ref{sec:MHV3pt} and \ref{sec:MHV4pt}, we will give explicit expressions for the tree-level color-ordered  three- and four-point amplitudes in the conformal basis.

\section{Gluon Three-Point Amplitudes}\label{sec:MHV3pt}

In this section we derive the  Mellin transforms of the  tree-level MHV and anti-MHV three-point amplitudes.   In the $(-+++)$ signature, the kinematics of massless scatterings forces the gluon three-point amplitude to vanish.  To circumvent this issue, we will instead  be working in the $(-+-+)$ signature in this section, in which $(z,\bz)$ as well as the spinor helicity variables $(|p],|p\>)$ are not related by complex conjugation, and the three-point function need not vanish.  In particular~(\ref{fga}) becomes:
\begin{align}\label{eq:mq}
q^\mu (z, \bar z)  = ( 1+z\bar z , z+\bar z , z-\bar z , 1-z \bar z)\,.
\end{align}
The Lorentz group in the $(-+-+)$ signature is $SL(2,\mathbb{R})\times SL(2,\mathbb{R})$, which acts on $z$ and $\bar z$ separately.

Let us start with the tree-level color-ordered MHV three-point amplitude.    
  Letting   the first and the second particles have negative helicities and the third particle have positive helicity,  the momentum space  amplitude is  (see Appendix \ref{app:convention} for our conventions on the spinor helicity variables),
\begin{align}
\mathcal{A}_{--+}(\omega_i ,z_i ,\bar z_i) &= {\< 12 \>^3 \over \<23\> \<31\>}\, \delta^{(4)} (p_1^\mu + p_2^\mu+p_3^\mu )\notag\\
&=-2{\omega_1 \omega_2 \over \omega_3 }{z_{12}^3 \over z_{23} z_{31}}\, 
\delta^{(4)} (\sum_i \varepsilon_i \omega_iq_i^\mu )\,.
\end{align}
The $2d$ spins of the corresponding conformal primaries are  $J_1=J_2 = -1$ and $J_3=+1$.  
In writing down the above expression, we have assumed $z_{ij}\neq0$, while $\bar z_{ij}$, which are independent real variables in the $(-+-+)$ spacetime signature, are allowed to vanish.  The sign   $\varepsilon_i$ is $+1$ ($-1$) for an outgoing (incoming) particle when we rotate back to the $(-+++)$ signature.  However, in the $(-+-+)$ signature, there is no invariant distinction between an incoming and an outgoing wavefunction. Indeed, under $SL(2,\mathbb{R})\times SL(2,\mathbb{R})$, the sign $\varepsilon_i$ is not invariant and transforms as $\varepsilon_i \to\varepsilon_i \text{sgn}((cz_i+d)(\bar c \bar z_i +\bar d))$, where $a,b,c,d,\bar a,\bar b, \bar c,\bar d\in\mathbb{R}$ with $ad-bc=\bar a\bar d-\bar b\bar c=1$.

We will follow the route of Section~\ref{sec:Mellin}. The delta function in \eqref{eq:mts} on the support of $z_{ij}\neq 0$ can be written
\begin{align}\label{eq:d1}
\left.\delta^{(4)}(\sum\limits_i\e_i\sigma_iq_i)\delta(\sum\limits_i\sigma_i-1)\right|_{z_{ij}\neq 0}
&= \frac{\delta(\bz_{12})\delta(\bz_{13})}{4\sigma_1\sigma_2\sigma_3  D_3^2}
\,
\delta(\sigma_1-\frac{z_{23}}{
D_3})\delta(\sigma_2+\e_1\e_2\frac{z_{13}}{ D_3})
\delta(\sigma_3-\e_1\e_3\frac{z_{12}}{D_3})\,,\notag\\
&\equiv\frac{\delta(\bz_{12})\delta(\bz_{13})}{4\sigma_1\sigma_2\sigma_3  D_3^2}
\,
\prod_{i=1}^3 \delta(\sigma_i -\sigma_{*i})\,,
\end{align}
where the denominator is
\begin{align}
D_3 = (1-{\e_1}{\e_2})z_{13}+({\e_1}{\e_3}-1)z_{12} \,.
\end{align}
  There is a similar term with support at $\bz_{ij}\neq 0$ relevant for the anti-MHV three-point amplitude, which by symmetry of the left hand side is just the above expression with the substitution $z_{ij}\leftrightarrow\bz_{ij}$.  Thanks  to the delta functions in \eqref{eq:d1}, all the Mellin integrals collapse to evaluating the integrand  on the solutions of $\sigma_i$:
\begin{align}\label{--+}
\boxed{\,\mathcal{ \tilde A}_{--+} (\lambda_i ;z_i,\bar z_i)  
= -
{\pi} \, \delta(\sum_i\lambda_i){\mathrm{sgn}(z_{12}z_{23}z_{31})\delta (\bar z_{13} ) \delta(\bar z_{12}) 
\over |z_{12}|^{-1- i \lambda_3}  |z_{23}| ^{1- i\lambda_1} |z_{13}| ^{1-i\lambda_2}}
\,\prod_{i=1}^3
\mathbf{1}_{[0,1]}(\sigma_{*i}),~{z_i,\bz_i\in\mathbb{R}}}\,
\end{align}
where $\sigma_{*i}$ are given in~(\ref{eq:d1}).     The indicator function $\mathbf{1}_{[0,1]}(x)$  is defined in \eqref{indicator}. Importantly, in the $(-+-+)$ signature, $z_i,\bar z_i$ are independent real variables and the notation $|z_{ij}|$ stands for the absolute value of a real variable, rather than $\sqrt{z_{ij}\bar z_{ij}}$.  Note that the sign function $\mathrm{sgn}(z_{12}z_{23}z_{31})$ is $SL(2,\mathbb{R})\times SL(2,\mathbb{R})$ invariant. 

Note that the three-point MHV amplitude  has a factor of $\frac{\sigma_i\sigma_j}{\sigma_k}$ for $i,j,k$ distinct.  This means that the denominator $D_3$ of $\sigma_{*i}$ drop out except for the indicator function constraints coming from the domain of integration of the simplex variables $\sigma_i$.  
In particular, the Mellin transform depends on the choice of the crossing channel (i.e. dependence on $\varepsilon_i$) \textit{only} through the  ranges of support for $z_i$'s constrained by the indicator functions $\prod_{i=1}^3
\mathbf{1}_{[0,1]}(\sigma_{*i})$.   

Let us decode the indicator functions $\prod_{i=1}^3
\mathbf{1}_{[0,1]}(\sigma_{*i})$. Their physical origin is that, given a crossing channel, not every possible direction, parametrized by $z_i,\bar z_i$, is allowed by the four-dimensional massless kinematics. For example,  the three-point function obviously vanishes if the three particles are all incoming or all outgoing. We therefore only need to consider the two-to-one or  one-to-two decay amplitudes, which will be denoted by  $ij\ce{<-->} k$, corresponding to $\e_i=\e_j=-\e_k$.  The two arrows of opposite directions are related by time reversal. The indicator functions $\prod_{i=1}^3
\mathbf{1}_{[0,1]}(\sigma_{*i})$ constrain the three real $z_i$'s to be in the following orderings for different crossing channels:
\be
\prod_{i=1}^3
\mathbf{1}_{[0,1]}(\sigma_{*i}):~~
\begin{array}{llll}
a)~~~ 12 \ce{<-->} 3~~~&\Rightarrow z_1<z_3<z_2  ~\mathrm{or}~ z_2<z_3<z_1&\\
b)~~~ 13 \ce{<-->} 2~~~&\Rightarrow z_1<z_2<z_3  ~\mathrm{or}~ z_3<z_2<z_1&\\
c)~~~ 23 \ce{<-->} 1~~~&\Rightarrow z_3<z_1<z_2  ~\mathrm{or}~ z_2<z_1<z_3&.\\
\end{array}
\ee
For each crossing channel, there are two possible orderings of the $z_i$'s.  Note that the ordering of two points $z_1$ and $z_2$ is not $SL(2,\mathbb{R})\times SL(2,\mathbb{R})$ invariant but depends on the sign of $(cz_1+d)(cz_2+d)$.  On the other hand, the crossing channel is also not invariant in the $(-+-+)$ signature.  The indicator functions are nonetheless $SL(2,\mathbb{R})\times SL(2,\mathbb{R})$ invariant if we take into account of the sign flip of $\varepsilon_i$ mentioned above, i.e. $\varepsilon_i \to\varepsilon_i \text{sgn}((cz_i+d)(\bar c \bar z_i +\bar d))$.

Coming back to the full three-point function, under an $SL(2,\mathbb{R})\times SL(2,\mathbb{R})$ action, the Mellin transform $\mathcal{ \tilde A} _{--+}(\lambda_i ;z_i,\bar z_i) $ of the color-ordered MHV amplitude indeed  transforms as a conformal three-point function of spin-one primaries with weights,
\begin{align}
\begin{split}
&h_1 = {i\over 2} \lambda_1 \,,~~~~~~~~~~~\bar h_1 = 1+ {i\over 2} \lambda_1\,,\\
&h_2 =  {i\over 2} \lambda_2 \,,~~~~~~~~~~~\bar h_2 = 1+ {i\over 2} \lambda_2\,,\\
&h_3 = 1+ {i\over2} \lambda_3 \,,~~~~~~\bar h_3  = {i\over 2}\lambda_3\,.
\end{split}
\end{align}
Note that it is important to use the conformal covariance of the delta function as in \eqref{deltacovariance}.

Next, consider the color-ordered anti-MHV amplitude where the first and second particles have positive helicities and the third particle has negative helicity,
\begin{align}
\mathcal{A}_{++-}(\omega_i ,z_i ,\bar z_i) &= {[ 12 ]^3 \over [23] [31]}\, \delta^{(4)} (p_1^\mu + p_2^\mu+p_3^\mu )\notag\\
&=2{\omega_1 \omega_2 \over \omega_3 }{\bz_{12}^3 \over \bz_{23} \bz_{31}}\, 
\delta^{(4)} (\sum_i \varepsilon_i \omega_iq_i^\mu )\,.
\end{align}
The $2d$ spins of the corresponding conformal primaries are  $J_1=J_2 = +1$ and $J_3=-1$.  
In writing down the above expression, we have assumed $\bz_{ij}\neq0$. 
Its Mellin transform is given by
\begin{align}\label{++-}
\boxed{\,
\mathcal{ \tilde A}_{++-} (\lambda_i ;z_i,\bar z_i)  
={\pi} \, \delta(\sum_i\lambda_i){\mathrm{sgn}(\bz_{12}\bz_{23}\bz_{31})\delta ( z_{13} ) \delta( z_{12}) 
\over  |\bar z_{12}|^{-1- i \lambda_3}  |\bar z_{23}| ^{1- i\lambda_1}|\bar z_{13} |^{1-i\lambda_2}}\,
\prod_{i=1}^3
\mathbf{1}_{[0,1]}(\sigma'_{*i}),~~~{z_i,\bz_i\in\mathbb{R}}}
\end{align}
where $\sigma'_{*i}$ are related to $\sigma_{*i}$ in~(\ref{eq:d1}) by $z_{ij}\leftrightarrow\bz_{ij}$.

Under an $SL(2,\mathbb{R})\times SL(2,\mathbb{R})$  action, $\mathcal{ \tilde A}_{++-} (\lambda_i ;z_i,\bar z_i) $ transforms as a conformal three-point function of spin-one primaries with weights,
\begin{align}
\begin{split}
&h_1 = 1+{i\over 2} \lambda_1 \,,~~~~~~\bar h_1 =  {i\over 2} \lambda_1\,,\\
&h_2 =  1+{i\over 2} \lambda_2 \,,~~~~~~\bar h_2 = {i\over 2} \lambda_2\,,\\
&h_3 = {i\over2} \lambda_3 \,,~~~~~~~~~~~\bar h_3  =1+ {i\over 2}\lambda_3\,.
\end{split}
\end{align}

\section{Gluon Four-Point Amplitudes}\label{sec:MHV4pt}
Let us move on to the tree-level color-ordered four-point MHV amplitude.  It is convenient to work in the $(-+++)$ spacetime signature in this section. We take  particles 1 and 2 to have negative helicities and particles 3 and 4 to have positive helicities.\footnote{Recall that we label the helicity of an external gluon as if it were an outgoing particle.}  We focus on the amplitude with  color order $(1234)$ \cite{Parke:1986gb},
\begin{align}
\mathcal{A} _{--++}(\omega_i,z_i,\bar z_i)= {\langle 12\rangle ^3 \over \langle 23\rangle \langle 34\rangle \langle 41\rangle}
\delta^{(4)} (  \sum_{i=1}^4 \varepsilon_i  \omega_i q_i )
\,,
\end{align}
where $\varepsilon_i$ is $+1$ ($-1$) if the particle is outgoing (incoming). 
The $2d$ spins of the corresponding conformal primaries are $J_1=J_2=-1$ and $J_3=J_4=+1$.

Again following the route of Section~\ref{sec:Mellin}, the four-point delta function in \eqref{eq:mts} can be written as
\begin{align}\label{eq:d2}
&\delta^{(4)}(\sum_{i=1}^4\e_i\sigma_iq_i)\delta(\sum_{i=1}^4\sigma_{i}-1)
= \frac{1}{4}\delta(|z_{12}z_{34}\bz_{13}\bz_{24}-z_{13}z_{24}\bz_{12}\bz_{34}|)
\notag\\
&\times\delta\left(\sigma_1+
\frac{\e_1\e_4}{ D_4 }
{z_{24}\bz_{34} \over z_{12}\bz_{13}}\right)
\delta\left(\sigma_2-
\frac{\e_2\e_4}{ D_4}
{z_{34}\bz_{14}\over z_{23}\bz_{12}}
\right)\delta\left(\sigma_3+\frac{\e_3\e_4}{ D_4}
{z_{24}\bz_{14}\over z_{23}\bz_{13}}\right)
\delta\left(\sigma_4-{1\over D_4}\right)  \notag\\
&\equiv \frac{1}{4}\delta(|z_{12}z_{34}\bz_{13}\bz_{24}-z_{13}z_{24}\bz_{12}\bz_{34}|)
\sum_{i=1}^4 \delta(\sigma_i  - \sigma_{*i})\,,
\end{align}
where the denominator $D_4$ is defined as
\begin{align}
D_4 = \left( 1-{\e_1\e_4}\right)
{z_{24} \bz_{34}\over z_{12}\bz_{13}}
+\left( {\e_2\e_4}-1\right)
{z_{34} \bz_{14} \over z_{23} \bz_{12}}
+\left(1-{\e_3\e_4}\right)
{z_{24}\bz_{14}\over z_{23}\bz_{13}} \,.
\end{align}
Note that due to the first delta function constraint, all of the $\sigma_{*i}$'s are real. 
The Mellin integrals in $\sigma_i$ all collapse to evaluating the integrand at $\sigma_i  = \sigma_{*i}$. 
By repeatedly using the first delta function constraint in \eqref{eq:d2}, we arrive at the following answer for the Mellin transform of the tree-level, color-ordered MHV four-point amplitude,
\begin{empheq}[box=\widefbox]{align}\label{eq:4ans} 
\mathcal{\tilde A} _{--++}(\lambda_i,z_i ,\bar z_i)=& -{\pi\over4}  \delta( \sum_k \lambda_k)  \delta\left({|z-\bar z|\over2}\right) \,
\left(  \prod_{i<j}^4 z_{ij}^{\frac h3 -h_i -h_j} \bar z_{ij}^{\frac {\bar h}{3} - \bar h_i -\bar h_j}\right)\,
 z^{\frac 53}\,(1-z)^{-\frac13}\, \notag\\
& \times\prod_{i=1}^4\mathbf{1}_{[0,1]}(\sigma_{*i})\,, \end{empheq}
where $\sigma_{*i}$ are given in~(\ref{eq:d2}) and the indicator function $\mathbf{1}_{[0,1]}(x)$ is defined in \eqref{indicator}. $z$ and $\bar z$ are the conformal cross ratios,
\begin{align}\label{crossratio}
z\equiv  {z_{12} z_{34} \over z_{13}z_{24}}\,,~~~~~~\bar z\equiv  {\bar z_{12} \bar z_{34} \over\bar z_{13}\bar z_{24}}\,,
\end{align}
and we have written our answer in terms of the weights
 \begin{align}\label{4weight}
\begin{split}
&h_1= {i\lambda_1\over2}\,,~~~~~~~~~h_2= {i\lambda_2\over2}\,,~~~~~~~~~\,
h_3= 1+{i\lambda_3\over2}\,,~~~~h_4= 1+{i\lambda_4\over2}\,,\\
&\bar h_1= 1+ {i\lambda_1\over2}\,,~~~~\bar h_2=1+ {i\lambda_2\over2}\,,~~~~
\bar h_3= {i\lambda_3\over2}\,,~~~~~~~~~\bar h_4= {i\lambda_4\over2}\,,
\end{split}
\end{align}
with $h\equiv \sum_{i=1}^4 h_ i$ and $\bar h  \equiv\sum_{i=1}^4 \bar h_i$.  
 As conventional in any CFT four-point function, we  write the answer as a product of a (non-unique) prefactor that accounts for the appropriate conformal covariance, and a function of the cross ratios which is conformal invariant.  Here, we chose this 
prefactor to be $ \prod_{i<j}^4 z_{ij}^{\frac h3 -h_i -h_j} \bar z_{ij}^{\frac {\bar h}{3} - \bar h_i -\bar h_j}$.  From~(\ref{eq:4ans}), we find that the Mellin transform of the color-ordered tree-level four-point amplitude does transform as a CFT four-point function with  the above weights.

From \eqref{eq:4ans}, we see that the delta function constrains the angular coordinates $z_i, \bar z_i$ on the celestial sphere such that the cross ratio $z$ is real,
\begin{align}\label{reality}
z-\bar z=0\,.
\end{align}
This constraint has a simple explanation. Using $SL(2,\mathbb{C})$ Lorentz invariance, we can arrange for the asymptotic positions of the first three particles $(z_1, z_2,z_3)$ to all lie on the equator of the celestial sphere. Momentum conservation then clearly implies the fourth must also lie on the equator. Interestingly, this is exactly the locus where a  Lorentzian CFT correlator in $(1+1)$  dimensions is singular as discussed in \cite{Maldacena:2015iua}.

Let us decode the indicator functions $\prod_{i=1}^4\mathbf{1}_{[0,1]}(\sigma_{*i})$.  The indicator functions is only non-vanishing for two-to-two scattering amplitudes, as the original amplitudes in momentum space vanish otherwise.  We will denote a two-to-two crossing channel as $ij \ce{<-->}k\ell$, corresponding to $\varepsilon_i =\varepsilon_j = - \varepsilon_k = -\varepsilon_\ell$.  From \eqref{eq:4ans}, we see that the four-point function depends on the crossing channel \textit{only} through the indicator functions $\prod_{i=1}^4\mathbf{1}_{[0,1]}(\sigma_{*i}) $, which  constrain the cross ratio to be in the following ranges for different crossing channels:
\be
\prod_{i=1}^4\mathbf{1}_{[0,1]}(\sigma_{*i}):~~\begin{array}{lll}
a)~~~ 12 \ce{<-->} 34~~~&\Rightarrow~~~ 1< z \\
b)~~~ 13 \ce{<-->} 24~~~&\Rightarrow~~~ 0<z<1\\
c)~~~ 14 \ce{<-->} 23~~~&\Rightarrow~~~  z<0\,,
\end{array}
\ee
in the $(- + ++)$ signature. Recall that the cross ratio is already constrained to be real $z=\bar z$ by the four-dimensional massless kinematics.  The indicator functions are manifestly $SL(2,\mathbb{C})$ invariant since it depends on the positions of the four points only through the cross ratios $z,\bar z$.

The MHV tree-level four-point amplitudes in other color orders can be obtained immediately by multiplying the answer~(\ref{eq:4ans})  by a function of $z_{ij}$.  For example,  the MHV amplitude in the color order $(1324)$ is given by
\begin{align}
  {\langle 12\rangle ^4 \over \langle 13\rangle \<32\>  \langle 24\rangle \langle 41\rangle}\delta^{(4)}(\sum_{i=1}^4 \varepsilon_i \omega_i q_i)
\,.
\end{align}
Its Mellin transform is simply~(\ref{eq:4ans}) multiplied by $-{z_{12} z_{34} \over z_{13}z_{24}}=-z$, so that the Mellin transform of the (1324) color-ordered amplitude again transforms as  a conformal four-point function with weights \eqref{4weight}, as expected.

\section{BCFW Recursion Relation of Conformal Correlators}\label{sec:BCFW}

In this section we transform the BCFW recursion  relation \cite{Britto:2004ap,Britto:2005fq} from  momentum space to the space of conformal primary wavefunctions. More explicitly, we perform Mellin transforms on both sides of the BCFW relation for the MHV four-point amplitude in terms of the three-point amplitudes.  We will be working in the $(-+-+)$ signature in this section where the three-point amplitude is  finite and well-defined.  Throughout this section, $|z_i|$ denotes the absolute value of a real variable, rather than $\sqrt{z_i\bar z_i}$.

\subsection{BCFW in Momentum Space}

Let us review the BCFW recursion  relation for the  four-point MHV amplitude.  We denote a stripped amplitude without the momentum conservation delta function as $A_n$, while the physical  unstripped amplitude is denoted as $\mathcal{A}_n$. 
The stripped color-ordered MHV and anti-MHV three-point amplitudes are
\begin{align}
A_3(1^-,2^-,3^+) = {\<12\>^3\over \<23\>\<31\>}\,,~~~~~
A_3 ( 1^+ ,2^+ ,3^- ) = {[12]^3\over [23][31]}\,.
\end{align}
On the other hand,  the stripped color-ordered MHV four-point amplitude is
\begin{align}
A_4(1^-,2^- ,3^+,4^+ )  = { \< 12\>^3 \over \<23\> \<34\> \<41\>} \,.
\end{align}
Let $p_i^\mu$ be the four-momenta of the four external particles satisfying the momentum conservation $\sum_{i=1}^4 p_i^\mu=0$.  Their spinor helicity variables will be denoted by $|i\>$ and $|i]$. We define $P^\mu_{i,j} \equiv p^\mu_i + p^\mu_j$.   To apply the BCFW recursion  relation, we choose 1 and 4 to be the reference gluons and shift their spinor helicity variables by
 \begin{align}\label{shift}
 \begin{split}
 &|\widehat 1\>  = |1\>\,,~~~\,~~~~~~~~~~~~ | \widehat 1 ]  = |1]  +u  |4]\,,\\
 &|\widehat 4\> = |4\> -u   |1\>\,,~~~~~~~|\widehat 4 ] = |4]\,,~~~~~~~~~~~~~
 \end{split}
 \end{align}
 where 
 \begin{align}
 &u \equiv - {P_{3,4}^2 \over \< 1 |P_{3,4} | 4]} \,,
 \end{align}
with $\< i | p_j |k ] = \< i j\> [jk]$.

 The BCFW recursion  relation equates the four-point amplitude to the product of two three-point amplitudes with shifted momenta,
 \begin{align}\label{bcfw4}
A_4(1^-,2^- ,3^+,4^+ ) = A_3 (\widehat1^- ,2^-, -\widehat P^+_{1,2}  ) {1\over P_{1,2}^2} 
 A_3( \widehat P^-_{1,2} ,3^+ ,\widehat 4^+).
 \end{align}
In the following section we will rewrite the above BCFW recursion relation for the four-point amplitude in the space of conformal primary wavefunctions and verify that it is obeyed by our expressions.

\subsection{BCFW of Conformal Correlators}

We  first want to determine the change in $\omega,z, \bar z$ under the BCFW shift.  We can choose a reference frame so that
\begin{align}
&|p\rangle =\pm \sqrt{2\omega} \left(\begin{array}{c}- 1 \\-z\end{array}\right)\,,~~~~~~|p]  = \sqrt{2\omega} \left(\begin{array}{c}-\bar z \\ 1\end{array}\right)\,,
\end{align}
where $p$ is a null vector that is parametrized by $\omega,z,\bar z$ as $p^\mu  = \pm\omega (  1+z\bar z , z+\bar z, z-\bar z, 1-z\bar z)$.  We choose a plus (minus) sign for an outgoing (incoming) momentum.  Let $\hat p$ be the BCFW-shifted momentum parametrized by $\widehat\omega, \widehat z, \widehat {\bar z}$.  Its spinor helicity variables are
\begin{align}\label{littlegroup}
&|\hat p\rangle =\pm t \,  \sqrt{2\widehat\omega} \left(\begin{array}{c}-1 \\ - \widehat z\end{array}\right)\,,~~~~~~|\hat p]  =  t^{-1}\, \sqrt{2\widehat\omega}\left(\begin{array}{c} -\widehat{\bar z} \\1\end{array}\right)\,.
\end{align}
Recall that given a null momentum, the spinor variables are only well-defined up to a little group rescaling $|p\> \to t|p\>,\, |p] \to t^{-1} |p]$ which will be important momentarily.
The BCFW shift \eqref{shift} for incoming particle number 1  is
\begin{align}
\begin{split}
&\sqrt{2\widehat \omega_1} \left(\begin{array}{c}1 \\ \widehat  z_1\end{array}\right)  = 
t_1^{-1}\sqrt{2\omega_1} \left(\begin{array}{c}1 \\z_1\end{array}\right)\,,\\
&\sqrt{2\widehat \omega_1} \left(\begin{array}{c}- \widehat  {\bar z}_1 \\ 1\end{array}\right)  = 
t_1  \left[ \,  \sqrt{2\omega_1} \left(\begin{array}{c}-\bar z_1 \\ 1\end{array}\right)
+u \sqrt{2\omega_4} \left(\begin{array}{c}-\bar z_4 \\ 1\end{array}\right)\,\right]
\,.
\end{split}
\end{align}
From above, we can express the hatted variables in terms of the unhatted ones,
\begin{align}
\begin{split}
&\widehat{\omega}_1= \omega_1  +  u  \sqrt{\omega_1\omega_4}   = {\bar z_{14}\over \bar z_{24} }\omega_1  \,,\\
&\widehat z_1 = z_1\,,\\
&\widehat{\bar z}_1= { \sqrt{\omega_1} \bar z_1 +u \sqrt{\omega_4} \bar z_4 
\over \sqrt{\omega_1} +  u \sqrt{\omega_4}  } = \bar z_2
\,.
\end{split}
\end{align}
The little group factor $t_1$ is given by $t_1=  \sqrt{\omega_1 \over  \widehat \omega_1} = \sqrt{ \bar z_{24} \over \bar z_{14}}$.  

Similarly, for outgoing particle number 4, the BCFW-shifted variables are
\begin{align}
\begin{split}
&\widehat{\omega}_4  = \omega_4  + u\sqrt{\omega_4\omega_1}  =  { z_{14}\over  z_{13} }\omega_4 \,,\\
&\widehat{ z}_4  = { \sqrt{\omega_4} z_4  + u \sqrt{\omega_1}  z_1 
\over \sqrt{\omega_4} + u \sqrt{\omega_1}  }   =z_3   \,,
\\
&\widehat {\bar z}_4 =\bar z_4\,.
\end{split}
\end{align}
The little group factor is given by $t_4 = \sqrt{\widehat \omega_4 \over \omega_4}  = \sqrt{  z_{14} \over z_{13} }$.  Particles 2 and 3 are unaffected by the BCFW shift so that $\widehat \omega_ i=\omega_i,\, \widehat z_i = z_ i  ,\, \widehat{\bar z}_i = \bar z_i$ for $i=2,3$.

The unshifted  (stripped) three-point amplitude $A_3(1^-,2^-,-P^+)$ can be written in terms of the $\omega, z,\bar z$ coordinates as follows
\begin{align}
A_3(1^-,2^-,-P^+)  =
{-} 2{\omega_1 \omega_2 \over \omega_{P}}{ z_{12}^3\over z_{2P} z_{P1}} \equiv 
A_{--+} (\omega_i ;z_i ,\bar z_i)\,.
\end{align}
For the BCFW-shifted three-point amplitude, we need to keep track of the little group factors in \eqref{littlegroup},
\begin{align}
A_3(\hat 1^-,2^-, - \widehat P^+)  =
{-}2 t_1^2 \,{\widehat\omega_1 \widehat\omega_2 \over \widehat\omega_P}{ \widehat z_{12}^3\over \widehat z_{2P}\widehat z_{P1}}
=
 t_1^2  \,A_{--+} (\widehat\omega_i ;\widehat z_i ,\widehat{\bar z}_i)\,.
\end{align}
Similarly, if
\begin{align}
A_3( P^- , 3^+ ,\hat 4^+)    =  2{\omega_3 \omega_4 \over\omega_P}{ {\bar  z}_{34}^3\over {\bar z}_{3P}{\bar z}_{P4}}
\equiv \,A_{-++} (\omega_i ; z_i ,{\bar z}_i)\,,
\end{align}
then
\begin{align}
A_3(\widehat P^- , 3^+ ,\hat 4^+)  = 2t_4^{-2} \,{\widehat\omega_3 \widehat\omega_4 \over \widehat\omega_P}{ \widehat {\bar  z}_{34}^3\over \widehat{\bar z}_{3P}\widehat {\bar z}_{P4}}
=
 t_4^{-2}  \,A_{-++} (\widehat\omega_i ;\widehat z_i ,\widehat{\bar z}_i)\,.
\end{align}

The BCFW relation is conventionally written in terms of the stripped amplitudes.  We would like to write it  in terms of the physical unstripped amplitudes, as only these have well-defined Mellin transforms.  We do not want to work directly with the standard BCFW-shifted variables $\widehat\omega_i ,\widehat z_i,  \widehat{\bar z}_i$ since the three-point function $\mathcal{\tilde A}_3$ is singular at  these values.    We will circumvent this difficulty by considering a modified shift of the variables $\omega_i , z_ i, \bar z_i$,
\begin{align}
\begin{split}\label{tilde}
& (\widetilde \omega_1 , \widetilde z_1 , \widetilde {\bar z}_1 ) = \left( |1+U \zeta | \omega_1 ,z_1 , {\bar z_1 + U \zeta \bar z_4 \over 1+ U \zeta}\right)\,,\\
&(\widetilde \omega_2  , \widetilde z_2 , \widetilde {\bar z}_2) =(\omega_2, z_2,\bar z_2)\,,\\
&(\widetilde \omega_3  , \widetilde z_3, \widetilde {\bar z}_3 ) = (\omega_3 ,z_3 , \bar z_3)\,,\\
&(\widetilde \omega_4  , \widetilde z_4, \widetilde {\bar z}_4 )  =
\left( |1+U \zeta^{-1}| \omega_4 ,{ z_4 +U\zeta^{-1} z_1 \over 1+ U\zeta^{-1}} ,\bar z_4\right)\,,
\end{split}
\end{align}
where $U$ is a free variable, the absolute values are such that $\widetilde \omega_i>0$, and
\begin{align}\label{eq:zeta}
\zeta\equiv \sqrt{ z_{12} \bar z_{13}  \over z_{24}  \bar z_{34}}\,.
\end{align}
Using~\eqref{eq:d2}, one can show that the quantity in the square root in~\eqref{eq:zeta} is positive.  The tilde variables $\widetilde \omega_i , \widetilde z_i$ coincide with the standard BCFW-shifted variables $|\widehat \omega_i|, \widehat z_i$ if both the ratio $\omega_4/\omega_1$ and $U$ equal to their on-shell values, i.e. if $ {\omega_4\over\omega_1 }= \zeta^2$ and $U=u$.  Indeed, the on-shell values of $\omega_1$ and $\omega_4$ can be computed using \eqref{eq:d2} to be
\begin{align}
{\omega_4 \over \omega_1} ={ z_{12}\bar z_{13}  \over z_{24} \bar z_{34} }  =\zeta^2\,.
\end{align}
Even though the two shifts agree on-shell, there is an important Jacobian factor  relating $\delta^{(4)}(p_1+p_2 + p_3 +p_4)$ and $\delta^{(4)}(\widetilde p_1 + p_2 +p_3 +\widetilde p_4)$:
\begin{align}\label{deltajacobian}
&\delta^{(4)}(p_1+p_2 + p_3 +p_4)=
\Big|{\text{det} \{\widetilde p_1 ,p_2 , p_3 , \widetilde p_4\} \over
  \text{det} \{p_1 ,p_2 , p_3 , p_4\}}\Big|
\delta^{(4)}(\widetilde p_1+ p_2 + p_3 + \widetilde p_4)\notag\\
&
\to_{U=u} |1-z| \, \delta^{(4)}(\widetilde p_1+ p_2 + p_3 +\widetilde p_4)\,.
\end{align}

Let us now rewrite the BCFW relation as
\begin{align}
&\mathcal{A}_4(1^-,2^-,3^+,4^+) 
= A_4(1^-,2^-,3^+,4^+) \delta^{(4)}(\widehat p_1 + p_2 +p_3 +\widehat p_4)\notag\\
& = t_1^2t_4^{-2} \,  {1\over P_{1,2}^2}  
 A_{--+} \left(\,  \widehat\omega_1, \omega_2 ,\widehat\omega_{P_{1,2}} ; \widehat z_i, \widehat {\bar z}_i\, \right)
     A_{-++} \left(\, \widehat\omega_{P_{1,2}} , \omega_3, \widehat\omega_4; \widehat z_i, \widehat {\bar z}_i\, \right)
 \notag\\
&\times\int d^4 P\, \delta^{(4)}(\widehat p_1 + p_2 - P) \delta^{(4)}( P+ p_3 +\widehat p_4)\,.
\end{align}
Next, we trivially insert $\int_{-\infty}^\infty dU \delta(U-u)$ into the integral and use
\begin{align}
{1\over P_{1,2}^2} \delta( U - u) = 
{-\mathrm{sgn}(z_{12}\bz_{12}){1\over |u|}} \delta( \<1|P_{1,2}|4] U- P_{1,2}^2) 
={-\mathrm{sgn}(z_{12}\bz_{12}){1\over |U|}} \delta(\widehat P_{1,2}^2(U)) \,,
\end{align}
where we have defined  $\widehat P_{1,2} (U)= P_{1,2} +U |4] \<1|$. At the on-shell value of $U$, i.e. $U=u$, $\widehat P_{1,2}(u) = \hat p_1+p_2$. Since the integrand only has support on $U=u$, we can replace $\widehat P_{1,2}(U)$ by $P$ and $\widehat \omega_i , \widehat z_i $ by $\widetilde\omega_i , \widetilde z_i $ (which depend on $U$).  The BCFW relation then takes the unstripped form
\begin{align}
&\mathcal{A}_4(1^-,2^-,3^+,4^+)  \notag\\
& =  {-\mathrm{sgn}(z_{12}\bz_{12})|1-z|}t_1^2t_4^{-2} \,    \int_{-\infty}^\infty {dU\over |U|}\,
\int d^4  P\,\delta(P^2)~
 A_{--+} \left(\,\widetilde\omega_1, \omega_2, \omega_{P}   ; \widetilde z_i , \widetilde{\bar z}_i\, \right)
\delta^{(4)}(\widetilde p_1 + p_2 - P) \notag\\
  &\times   
  A_{-++} \left(\, \omega_{P}, \omega_3,\widetilde\omega_4  ; \widetilde z_i, \widetilde {\bar z}_i\, \right)
\delta^{(4)}( P+ p_3 +\widetilde p_4)  \\
& = {-\mathrm{sgn}(z_{12}\bz_{12})|1-z|} t_1^2t_4^{-2} \,   \int_{-\infty}^\infty {dU\over |U|}\,
\int d^4 P \,  \delta ( P^2)  \,
 \mathcal{A}_{--+} \left(\,  \widetilde\omega_1, \omega_2,\omega_P  ; \widetilde z_i, \widetilde {\bar z}_i\, \right)
\mathcal{A}_{-++} \left(\, \omega_P,\omega_3, \widetilde\omega_4   ; \widetilde z_i, \widetilde {\bar z}_i\, \right) \, ,\notag
\end{align}
where $\mathcal{A}$ denotes the unstripped amplitude with the momentum conservation delta function included. The Jacobian factor ${|1-z|}$ comes from the momentum conservation delta functions as explained in \eqref{deltajacobian}.  
  Because of the delta function $\delta( P^2)$, $P^\mu$ is null and we can define $ \omega_P, \,  z_P, \,  {\bar z}_P$  as $P^\mu  =  \omega_P  \, (1+z_P\bar z_P ,  z_P+{\bar z}_P , z_P - {\bar z}_P , 1-z_P\bar z_P)$. 
  In $\mathcal{A}_{--+}$ above and from now on, the notation $\widetilde z_i $ collectively denotes $\widetilde z_3 , \, \widetilde z_4,$ and $z_P$. We use a similar collective notation in $\mathcal{A}_{-++}$.

 We now perform  the Mellin transform on both sides of the BCFW relation,
  \begin{align}
&\mathcal{\tilde A}_{--++}(\lambda_i , z_i ,\bar z_i ) 
 \equiv \prod_{i=1}^4 \int_{0}^\infty d\omega_i  \, \omega_i^{i\lambda_i}  \,  \mathcal{A}_4(1^-,2^-,3^+,4^+)\notag\\
&= {-\mathrm{sgn}(z_{12}\bz_{12})|1-z| \left| {\bar z_{24}\over \bar z_{14}}\right|^{2+i\lambda_1} \left| {z_{13}\over z_{14}}\right|^{2+i\lambda_4}}
  \int_{-\infty}^\infty {dU\over |U|}\,
  \int d^4 P \delta(P^2)  \prod_{i=1}^4 \int_{0}^\infty d\widetilde\omega_i \, \widetilde\omega_i^{i\lambda_i}\, \notag\\
&\times
 \mathcal{A}_{--+} \left(\,  \widetilde\omega_1 , \omega_2 ,  \omega_P; \widetilde z_i, \widetilde {\bar z}_i\, \right)
\mathcal{A}_{-++} \left(\, \omega_P ,  \omega_3, \widetilde\omega_4 ; \widetilde z_i, \widetilde {\bar z}_i\, \right) 
\, \notag\\
&= {-\mathrm{sgn}(z_{12}\bz_{12})|1-z|\left| {\bar z_{24}\over \bar z_{14}}\right|^{2+i\lambda_1} \left| {z_{13}\over z_{14}}\right|^{2+i\lambda_4}}
\int_{-\infty}^\infty {dU\over |U|}   \,     
\int { d^4 P}\delta(P^2)   \prod_{i=1}^4 \int_{0}^\infty d\widetilde\omega_i  \, \widetilde\omega_i^{i\lambda_i} \,\notag\\
&\times
\prod_{i=1}^4 \int_{-\infty}^\infty {d\widetilde \lambda_i \over 2\pi} 
\int_{-\infty}^\infty  {d\lambda_P \over 2\pi} 
\int_{-\infty}^\infty  {d\lambda_{P'} \over 2\pi} 
\left(\prod_{i=1}^4\widetilde\omega_i^{-1-i\widetilde\lambda_i}     \right)
\omega_P^{-2-i\lambda_P-i \lambda_{P'}} \,\notag\\
&\times   
\mathcal{\tilde A}_{--+}( \widetilde\lambda_1, \widetilde\lambda_2 , \lambda_P;  \widetilde z_j , \widetilde{\bar z}_j)
\mathcal{\tilde A}_{-++}( \lambda_{P'},  \widetilde\lambda_3 ,\widetilde\lambda_4; \widetilde z_j , \widetilde{\bar z}_j)\,.
\end{align}
In the second line we have changed the integration variables to  the shifted energy $\widetilde{\omega}_i$.  In the third line we express the three-point amplitudes in terms of their Mellin transforms. 

 Finally, using
\begin{align}
\int { d^4 P}\delta(P^2)  = 
 \int_0^\infty  \, \omega_P d \omega_P
  \, \int dz_P d {\bar z}_P\, ,  
\end{align}
we can perform the $ \omega_P$ and the $\lambda_{P'}$ integrals to obtain,
\begin{align}
&\mathcal{\tilde A}_{--++}(\lambda_i , z_i ,\bar z_i ) 
 ={-\mathrm{sgn}(z_{12}\bz_{12})|1-z|\left| {\bar z_{24}\over \bar z_{14}}\right|^{2+i\lambda_1} \left| {z_{13}\over z_{14}}\right|^{2+i\lambda_4}}
   \,\notag\\
&~\times
\int_{-\infty}^\infty {dU\over |U|}   \,     
   \int_{-\infty}^\infty  {d \lambda_P \over 2\pi}  
   \int d z_P d{\bar z}_P
\mathcal{\tilde A}_{--+}( \lambda_1,\lambda_2 , \lambda_P;  \widetilde z_j , \widetilde{\bar z}_j)
\mathcal{\tilde A}_{-++}( -\lambda_{P}, \lambda_3, \lambda_4 ; \widetilde z_j , \widetilde{\bar z}_j)
\,.
\label{BCFWMellin}
\end{align}
The above equation is our final result for the BCFW recursion  relation in the space of conformal primary wavefunctions. The tilde variables are defined in \eqref{tilde} and $\widetilde z_P=z_P$, $\widetilde {\bar z}_P=\bar z_P$.

~\\

Let us check \eqref{BCFWMellin} by explicitly plugging in the three-point functions obtained in \eqref{--+} and \eqref{++-}:
\begin{align}
&\mathcal{\tilde A}_{--+}(\lambda_1,\lambda_2 , \lambda_P;  \widetilde z_j , \widetilde{\bar z}_j)
=-
{\pi} \mathrm{sgn}(z_{12}z_{2P}z_{P1})  \, \delta(\lambda_1+\lambda_2+\lambda_P){\delta (\widetilde{\bar z}_{1} - \widetilde{ \bar z}_2 )
 \delta(\widetilde{\bar z}_{2P} ) 
\over  |z_{12}|^{-1- i \lambda_P}  |z_{2P}|^{1- i\lambda_1} |z_{1P}| ^{1-i\lambda_2}}    \,,\notag\\
&\mathcal{\tilde A}_{-++}(-\lambda_{P},\lambda_3 , \lambda_4, ; \widetilde z_j, \widetilde{\bar z}_j)
={\pi} \mathrm{sgn}({\bz_{34}\bz_{4P}\bz_{P3}}) \, \delta(\lambda_3+\lambda_4-\lambda_P){\delta (\widetilde{ z}_{4} - \widetilde{  z}_3 )
 \delta(\widetilde{ z}_{3P} ) 
\over  |\bar z_{43}|^{-1+i \lambda_P}  |\bar z_{3P}| ^{1- i\lambda_4} |\bar z_{4P}| ^{1-i\lambda_3}}   \,.
\end{align}
All the integrals can be performed by solving the delta functions. In particular we have
\begin{align}
U= {z_{34} \over z_{13}}  \zeta\,. 
\end{align}
Plugging the above value of $U$ into another delta function we obtain
\begin{align}
&\delta\left({\bar z_{12} - U \zeta\bar z_{24}   
\over 1+U \zeta  }\right)
= \delta(|z-\bar z|) \, \left|{ \bar z_{34} \bar z_{14}\over\bar z_{24}^2 \bar z_{13}}\right|\,,
\end{align}
where $z$ and $\bar z$ are the cross-ratios \eqref{crossratio}. One can then verify that the righthand side of \eqref{BCFWMellin} correctly reproduces the $(-+-+)$ signature analog of the four-point function \eqref{eq:4ans}.\footnote{There is a $\mathrm{sgn}(z_{ij}\bz_{ij})$ in the $(-+-+)$ analog of \eqref{eq:4ans} which is matched by the signs in the BCFW recursion relation.}

~\\~\\~\\
\section*{Acknowledgements}
We are  grateful to N. Arkani-Hamed, C. Cheung, D. Harlow, Y.-t. Huang,  D. Jafferis,  J. Maldacena, P. Mitra,  B. Schwab,  and A. Zhiboedov for useful conversations.   This work was supported in part by DOE grant DE-SC0007870.  S.P. is supported by the National Science Foundation and by the Hertz Foundation through a Harold and Ruth Newman Fellowship.  S.H.S. is supported by the National Science Foundation grant PHY-1606531.
~\\~\\
\appendix
\section{Conventions}\label{app:convention}

In this appendix we will review our conventions in the $\eta_{\mu\nu}=\text{diag}(-1,+1,+1,+1)$ signature. The Levi-Civita symbols are normalized as $\epsilon^{12} = -\epsilon^{21}= +1$ and $\epsilon_{12}=-\epsilon_{21} =-1$.  We denote the chiral and anti-chiral spinor indices of the Lorentz group $SL(2,\mathbb{C})$ as $\alpha$ and $\dot\alpha$, respectively. The index of a spinor $\lambda_\alpha$ is lowered and raised as $\lambda^\alpha = \epsilon^{\alpha\beta} \lambda_\beta$ and $\lambda_\alpha = \epsilon_{\alpha\beta} \lambda^\beta$.

A four-momentum $p^\mu$ can be represented by a two-by-two matrix as
\begin{align}
p_{\alpha\dot\alpha} = p_\mu \sigma^\mu_{\alpha\dot\alpha}\,,
\end{align}
where $\sigma^\mu_{\alpha\dot\alpha} = (I, \vec \sigma)$.  We also define
\begin{align}
p^{\dot\alpha\alpha}  = \epsilon^{\dot\alpha\dot\beta}\epsilon^{\alpha\beta} p_{\beta\dot\beta} 
= p_\mu \bar\sigma^{\mu\dot\alpha\alpha}\,,
\end{align}
where $\bar \sigma^{\mu\dot\alpha\alpha} = (I,-\vec\sigma)$.  Using the identity $\epsilon^{\alpha\beta} \epsilon^{\dot\alpha\dot\beta} \sigma^\mu _{\alpha\dot\alpha} \sigma^\nu_{\beta\dot\beta} = -2\eta^{\mu\nu}$, the inner product between two four-momenta can be written as $\epsilon^{\alpha\beta}\epsilon^{\dot\alpha\dot\beta} p_{\alpha\dot\alpha} q_{\beta\dot\beta} 
=-2p^\mu q_\mu\,.$

A null four-momentum $p_{\alpha \dot\alpha}$ has vanishing determinant so can always be written in terms of their spinor helicity variables,
\begin{align}
p_{\alpha\dot\alpha} =  |p] _\alpha \, \< p|_{\dot \alpha}\,.
\end{align}
The spinor helicity variables are defined up to a little group rescaling, $|p\>\to t|p\>$ and $|p] \to t^{-1}|p]$.
Similarly,
\begin{align}
p^{\dot\alpha\alpha} = | p \>^{\dot\alpha} \, [ p|^\alpha\,,
\end{align}
where the spinor helicity variables with upper indices are defined as
\begin{align}
[p|^\alpha = \epsilon^{\alpha\beta} |p]_\beta\,,~~~~~~\< p|_{\dot \alpha}  = \epsilon_{\dot\alpha\dot\beta} |p\>^{\dot\beta}\,.
\end{align}
We  define the brackets of spinor helicity variables as
\begin{align}
&[pq] = [ p|^\alpha \, |q]_\alpha =- \epsilon^{\alpha\beta} |p]_\alpha \, |q]_\beta  = -[qp]\,,\\
&\<pq\> = \<p|_{\dot\alpha}  \,|q\>^{\dot\alpha} =  \epsilon^{\dot\alpha\dot\beta} \<p|_{\dot\alpha}\<q| _{\dot\beta} = -\<qp\>\,.
\end{align}
The inner product between two null momenta can be written as products of the brackets,
\begin{align}
2 p\cdot q = - p_{\alpha \dot\alpha} q_{\beta\dot\beta} \epsilon^{\alpha\beta}\epsilon^{\dot\alpha\dot\beta}= \<pq\> [pq]\,.
\end{align}
We also define:
\begin{align}
&[p| k | q\>  =  [p|^\alpha \, k_{\alpha\dot\alpha } \, |q\>^{\dot\alpha} = [ p k] \<kq\>\,,\\
&\< q | k |p] = \<q |_{\dot\alpha} \, k^{\dot\alpha\alpha}  \, |p] _{\alpha}  =  \<qk\>[kp]\,.
\end{align}

We can choose a frame and parametrize a null momentum $p^\mu$ by $\omega, z, \bar z$ as,
\begin{align}
p^\mu  = \pm\omega( 1+|z|^2 , z+\bar z ,  -i (z-\bar z) ,  1-|z|^2)\,,
\end{align}
with a plus (minus) sign for an outgoing (incoming) momentum.  
In terms of a two-by-two matrix, we have
\begin{align}
p_{\alpha\dot \alpha} = \sigma^\mu_{\alpha\dot \alpha} p_\mu = |p]_\alpha \< p|_{\dot \alpha} = \pm 2\omega\left(\begin{array}{cc}-|z|^2 & \bar z \\z & ~-1\end{array}\right)\,.
\end{align}
Next we want to express the spinor helicity variables  in terms of $\omega,z,\bar z$.  A priori, any such  identification suffers from the   ambiguity of little group rescaling $|p\>\to t|p\>, |p]\to t^{-1} |p]$, which in turn rescales the polarization vectors as $\epsilon^\mu_\pm(p) \to t^{\mp 2} \epsilon^\mu_\pm(p)$.  
For our purpose, however,  the choice of conformal primary wavefunction in \eqref{gaugerep} fixes a particular normalization for the polarization vectors as in \eqref{polarization}, $\epsilon_+^\mu(p=\pm\omega q )  = {1\over \sqrt{2}}\partial_z q^\mu$ and $\epsilon^\mu_-(p=\pm\omega q) ={1\over \sqrt{2}}\partial_\bz q^\mu$. In this normalization the spinor helicity variables can be written in terms of $\omega,z,\bar z$ as 
\begin{align}
\begin{split}
&|p]_\alpha = \sqrt{2\omega}\left(\begin{array}{c}-\bar z \\ 1\end{array}\right)\,,~~~~~
\<p|_{\dot\alpha }= \pm\sqrt{2\omega}\left(\begin{array}{c}z \\ -1\end{array}\right)\,,\\
&[p|^\alpha = \sqrt{2\omega}\left(\begin{array}{c} 1 \\ \bar z\end{array}\right)\,,~~~~~~~
|p\>^{\dot\alpha }=\pm \sqrt{2\omega}\left(\begin{array}{c}-1 \\ -z\end{array}\right)\,.
\end{split}
\end{align}
For two outgoing or two incoming particles, the brackets can be written as
\begin{align}
[ij] =  2\sqrt{\omega_i \omega_j} (\bar z_i -\bar z_j)\,,~~~~~~\<ij\> = -2\sqrt{\omega_i \omega_j } ( z_i - z_j)\,.
\end{align}
On the other hand, the brackets between one incoming and one outgoing particle are
\begin{align}
[ij] =  2\sqrt{\omega_i \omega_j} (\bar z_i -\bar z_j)\,,~~~~~~\<ij\> = 2\sqrt{\omega_i \omega_j } (z_i - z_j)\,.
\end{align}

\section{Inner Products of One-Particle States}\label{sec:2pt}

In \cite{Pasterski:2017kqt} (see also \cite{deBoer:2003vf,Pasterski:2016qvg} for the scalar case) the inner product between four-dimensional  one-particle states of spin one was computed in the space of conformal primary wavefunctions. Here we review this calculation for completeness.  Let us denote a massless one-particle state with helicity $\ell=\pm1$ and three-momentum $\vec p$ by
$| \vec p, \ell\> $,
with the energy $p^0$ given by $p^0 = | \vec p|$.  The inner product between such one-particle states is\footnote{As before, we label a helicity of a gauge boson as it were an outgoing particle. That is why the inner product is only non-vanishing if $\ell_1= -\ell_2$.}
\begin{align}
\< p_2,\ell_2|  p_1 ,\ell_1 \>  =  2p_1^0(2\pi)^3 \, \delta_{\ell_1, -\ell_2}  \,\delta^{(3)} (\vec p_1+\vec p_2)\,.
\end{align}
The Mellin transform of this inner product is
\begin{align}
&\mathcal{\tilde A}_{J_1J_2}(\lambda_i,z_i,\bar z_i)   =(2\pi)^3 \,\delta_{\ell_1, - \ell_2}\,\notag\\
&\times \int_0^\infty d\omega_1 \,\omega_1^{i\lambda_1} \, 
\int_0^\infty  d\omega_2 \, \omega_2^{\,i\lambda_2} \,
\omega_1 (1+|z_1|^2) \, \delta^{(2) }( \omega_1 z_1- \omega_2 z_2 )
\delta(\omega_1 (1-|z_1|^2 ) -\omega_2 (1-|z_2|^2))\notag\\
&= (2\pi)^4\, \delta_{\ell_1,-\ell_2} \, \delta(\lambda_1+ \lambda_2)
\delta^{(2)} (z_1-z_2)\,,
\end{align}
where we have used \eqref{mellindelta}. 
The $2d$ spins are given by  $J_1=\ell_1$ and $J_2=\ell_2$.

Let us consider the case with helicities $-\ell_1=\ell_2=+1$, while the other case follows similarly.  The answer $\mathcal{\tilde A}_{-+}(\lambda_i,z_i,\bar z_i)$ is a contact term, but it has the same $SL(2,\mathbb{C})$ covariance as a two-point function
 of conformal primaries  with  weights (with $\lambda_2=-\lambda_1$ fixed by the delta function above)
\begin{align}
&h_1=i {\lambda_1\over2}\,,~~~~~~~~~\,~~ \bar h_1 = 1+i  { \lambda_1\over2}\,,\notag\\
&h_2 = 1+ i  {\lambda_2\over2}\,,~~~~~~\bar h_2 = i {\lambda_2\over2}\,.
\end{align}
In particular, the $2d$ spins are $J_1={h_1-\bar h_1} = -1$ and $J_2= {h_2-\bar h_2}=+1$.  Indeed, under an $SL(2,\mathbb{C})$ transformation $z_i\to z_i'  = {az_i +b\over cz_ i +d}$, the contact term $\delta(\lambda_1+\lambda_2)\delta^{(2)}(z_1-z_2)$ transforms as
\begin{align}\label{deltacovariance}
\delta(\lambda_1+\lambda_2)\delta^{(2) }(z_1' -z_2' )  &= |  cz_1+d |^{4}\delta(\lambda_1+\lambda_2)  \delta^{(2)}(z_1-z_2)\notag\\
&=
\left[ \prod_{i=1}^2 (cz_i+d)^{\Delta_i+J_i} (\bar c\bar z_i +\bar d)^{\Delta_i-J_i}  \right]\, \delta(\lambda_1+\lambda_2)  \delta^{(2)}(z_1-z_2)\,.
\end{align}

\bibliography{mellinfinal}{}
\bibliographystyle{utphys}

\end{document}